\documentclass{PoS}
\usepackage{mathtools}

\title{Neutrino quantum decoherence due to entanglement with magnetic field}

\ShortTitle{Neutrino quantum decoherence due to entanglement with magnetic field}

\author{Konstantin Stankevich \\
       Department of Theoretical Physics, Moscow State University, 119992 Moscow, Russia\\
       E-mail: \email{kl.stankevich@physics.msu.ru}}
\author{\speaker{Alexander Studenikin} \\
       Department of Theoretical Physics, Moscow State University, 119992 Moscow, Russia\\
       Dzhelepov Laboratory of Nuclear Problems, Joint Institute for Nuclear Research, 141980 Dubna, Russia \\
       E-mail: \email{studenik@srd.sinp.msu.ru}}

\abstract{The phenomena of neutrino oscillations emerges due to coherent superposition of different neutrino states. The entanglement of neutrinos with its environment can lead to a suppression of neutrino oscillations. The master equation for neutrino evolution in a magnetic field is derived taking into account the entanglement with a magnetic field.

}

\FullConference{EPS-HEP 2017, European Physical Society conference on High Energy Physics\\
		5-12 July 2017\\
		Venice, Italy}

\begin{document}

\section{Introduction}
It is well known that an external environment can significantly influence on neutrino oscillations pattern. In  studies of neutrino evolution in a magnetic field it is usually considered neutrinos as a closed system (see for instance \cite{Giunti-Stud-RMP:2015}). That means that the neutrinos do not have any backforward influence on a magnetic field. This influence can be negligible, in principle, but it causes entanglement between neutrinos and the environment. The entanglement can destroy quantum coherent superposition of different neutrino states and thus it can lead to the suppression of neutrino oscillations.

There are three types of quantum decoherence discussing in the literature: 1) dephasing, 2) entanglement with  the enviroment and 3) relevation of "which-path" information \cite{Hsiang-Ford:2008}. There is another type of decoherence in neutrino physics that is due to wave neutrino separation which is well studied in the literature and can be referred to a classic decoherence. This last phenomena we do not consider in this paper.

   The Lindblad's equation is often used in studies of quantum decoherence of neutrinos (see for example \cite{Nogueira-Souza:2017} and \cite{Gomes-Guzzo:2017}). However, there is a serious unsolved inherent problem of this approach that is manifested by a fact that the decoherence parameter is not fixed by the theory and remaining free. This provides an uncertainty or undesired freedom, while the prediction of the theory is compared with an experimental data. The aim of the present note is to fill the mentioned about gap and is dedicated to derivation of the equation for neutrino density matrix accounting for the effects of quantum decoherence due to entanglement with a magnetic field.

\section{General formalism}
Evolution of a closed system (which is neutrino plus magnetic field) in our case is governed by the Liouville equation

 \begin{equation}
i \frac{ \partial \rho (t)}{\partial t} = \left[H(t),\rho(t) \right]
\label{fonN}
,\end{equation}
where $\rho(t)$ is a density matrix that can describe a pure or a mixed state of the system. If we consider the system composed of two subsystems corresponding to neutrinos and the background magnetic field the hamiltonian is composed of three terms:

\begin{equation}
H = H_\nu+H_A+H_{int}
, \end{equation}
where $H_{int}$ stands for neutrinos to magnetic field interactions. If not to be interested in the evolution of environment (the magnetic field), then its degrees of freedom can be traced out and from \ref{fonN} we get

 \begin{equation}
i \frac{ \partial \rho_\nu (t)}{\partial t} = \left[H_\nu(t),\rho_\nu(t)\right]  + Tr_A\left(H_{int}\rho - \rho H_{int}\right)
\label{close}
, \end{equation}
where $\rho_\nu = Tr_A  \rho$ is a density matrix which describes the evolution of the studied neutrino system. Note that tracing out makes the equation irreversible.

\section{Master equation for neutrino}
A neutrino can interact with a magnetic field due to the anomalous magnetic moment $\mu$. The derivation of the evolution equation below is based on the analogous derivation in the case of particle interaction with a magnetic field due to an electric charge \cite{Petruccione}. We will employ the interaction representation. The interaction Hamiltonian is given by

\begin{equation}
H_{int} = \mu \bar{\psi}(x)\sigma_{\mu \nu}\psi(x) F^{\mu \nu}(x) = j_{\mu \nu}  F^{\mu \nu}
,\end{equation}
where $F^{\mu \nu} = \partial^{\mu} A^{\nu} - \partial^{\nu} A^{\mu}$ is an electromagnetic field tensor, $\psi(x)$ is a wave function of neutrino in the mass basis and

\begin{equation}
\mu \bar{\psi}(x)\sigma_{\mu \nu}\psi(x) = j_{\mu \nu},
.\end{equation}

 The field $A_\mu$ can be written in the form $A_{\mu}  = \bar{A}_\mu + a_\mu,$ where $\bar{A}_\mu$ describes a classic field $B$ and $a_\mu$ describes electromagnetic fluctuations. Here below we focus on the entanglement of the neutrino with the magnetic field. Therefore, for simplicity we omit $\bar{A}_\mu$ part in further calculations from here further on.


Equation (\ref{close}) can be formally solved (integrated). Then after tracing out electromagnetic degrees of freedom we obtain the following equation for the neutrino density matrix in mass basis:

\begin{equation}
\rho_{\nu} (t_f) = tr_A \left[ T exp \left( \int^{t_f}_{t_i} d^4xL(x) \right) \rho(t_i) \right]
,
\label{ExpInit}\end{equation}
where we introduce the Liouville superoperator $L(x)\rho_{\nu} = - i \left[  H_{int},\rho_{\nu} \right] \label{a1}$. Using the Wick's theorem \cite{Itzykson_Zuber}:

\begin{equation}
T^A \exp \left[\int^{t_f}_{t_i}d^4xL(x)\right] = \exp \left[\frac{1}{2} \int^{t_f} d^4x \int^{t_f} d^4x'  [L(x), L(x')] \theta(t-t') \right] \exp \left[\int^{t_f} d^4x L(x)\right]
,\end{equation}
we can exclude path-ordering ($T_A$) of an electromagnetic field and get

 \begin{multline}
\rho_{\nu}(t_f) =
T_{\nu} \left( -\frac{1}{2} \right. \exp \left[ \int^{t_f}_{t_i}d^4x \int^{t_f}_{t_i}d^4x'  \Theta (t-t')[F_{\mu\nu}(x),F_{\alpha\beta}(x')]  J_+^{\mu\nu}(x) J_+^{\alpha\beta}(x') \right. - \\
- \left. \left. \int^{t_f}_{t_i}d^4x \int^{t_f}_{t_i}d^4x'  \Theta (t-t') J_-^{\mu\nu}(x)  J_-^{\alpha\beta}(x')  \right] \right)
 \times tr_A \left( \exp \left[ \int^{t_f}_{t_i}d^4x L(x) \right] \right)
,
\label{afterT}
\end{multline}
where $T_\nu$ is path-ordering of the neutrino degrees of freedom and we use the following notations

\begin{equation}
J_+^{\mu\nu} (x)\rho_{\nu} = j^{\mu\nu}(x) \rho_{\nu} ,\;\;\; J_-^{\mu\nu}(x) \rho_{\nu} = \rho_{\nu} j^{\mu\nu}(x)
,\end{equation}

One can see that in equation (\ref{afterT}) averaging over magnetic field  degrees of freedom is included only in the functional

\begin{equation}
 W[J_+,J_-] = tr_A \left( \exp \left[ \int^{t_f}_{t_i}d^4x L(x) \right] \right)
.\end{equation}

To simplify this functional it is convenient to use a cumulative decomposition. Under the initial conditions of weak entanglement of the neutrinos with an external field ( $\rho = \rho_\nu \oplus \rho_A$ ), the maximum degree of the cumulative decomposition is two. Moreover,$<a_\mu(x)> = 0$, thus there is only one term in the cumulative decomposition. In the light of the above, functional $W[J_+,J_-]$ is written in the form

\begin{equation}
tr_A \left(\exp \left[ \int^{t_f}_{t_i}d^4x L(x) \right] \right)  = \exp \left[\frac{1}{2} \int^{t_f} d^4x \int^{t_f} d^4x' <L(x)L(x')> \right]
\label{Tracing}
.\end{equation}

With using (\ref{Tracing}) equation (\ref{afterT}) is modifyed as

\begin{equation}
\rho_{\nu}(t_f) = T^\nu \exp (i \Phi[J_+, J_-])\rho_{\nu}(t_i),
\end{equation}
where
\begin{multline}
i \Phi[J_+, J_-] = \frac{1}{2} \int^{t_f}_{t_i} d^4x \int^{t_f}_{t_i} d^4x' \times\\
\times [-i \partial_\alpha \partial_\mu D_F(x-x')_{\beta\nu}J_+^{\alpha\beta}(x) J_+^{\mu\nu}(x') + \partial_\alpha \partial_\mu D^*_F(x-x')_{\beta\nu}J_-^{\alpha\beta}(x) J_-^{\mu\nu}(x') + \\
+ \partial_\alpha \partial_\mu D_-(x-x')_{\beta\nu}J_+^{\alpha\beta}(x) J_-^{\mu\nu}(x') + \partial_\alpha \partial_\mu D_+(x-x')_{\beta\nu}J_-^{\alpha\beta}(x) J_+^{\mu\nu}(x')]
.\end{multline}
where $D_F(x-x')_{\mu\nu}$ is a Feynman propagator, $D_+(x-x')_{\mu\nu}$ and $D_-(x-x')_{\mu\nu}$ are two-point correlation functions. Further it will be also convenient to use the anticommutator function $D_1(x-x')_{\mu\nu}$:

\begin{gather}
D_F(x-x')_{\mu\nu} = <T[a_\mu(x),a_\nu(x')]>,\\
D_+(x-x')_{\mu\nu} = <T[a_\mu(x),a_\nu(x')]> ,\\
D_+(x-x')_{\mu\nu} = <T[a_\nu(x'),a_\mu(x)]>,\\
D_1(x-x')_{\mu\nu} = <T \{ a_\mu(x),a_\nu(x')\}>
.\end{gather}

In the following we use the Feynman gauge $D(x-x')_{ij} = \delta_{ij}D(x-x')$ with $a_\mu = (0,-\vec{a})$. The currents $j_{ij}$ can be expressed as

\begin{equation}
j_{ij} = 2 \mu \bar{\nu}(x)\gamma_0\gamma_i \gamma_j \nu(x) = 2 \epsilon_{ijk} S^k
,\end{equation}
where we introduced the spin vector

\begin{equation}
S^k = \mu \bar{\nu}(x)\gamma_0 \Sigma^k \nu(x)
.\end{equation}

Finally we arrive to a final formula for a density neutrino matrix

\begin{equation}
\frac{d}{dt} \rho_{\nu} (t) = K_0 \rho_{\nu}(t) + K_1 \rho_{\nu}(t) + K_2 \rho_{\nu}(t)
,\label{A3Final}
\end{equation}
where the superoperators are defined as

\begin{gather}
K_0 \rho_{\nu} = -\frac{1}{i} \left[ H_B, \rho_{\nu} \right]\\
K_1 \rho_{\nu} = - \frac{1}{2} \int d^3x \int d^3x' \int dx_0' \vec{\partial}^2D_1(x-x') [\vec{S} (x),  [\vec{S} (x'), \rho_{\nu} ] ], \\
K_2 \rho_{\nu} =  \sum_{i,j} \frac{1}{2} \int d^3x \int d^3x' \int dx_0' \partial_i\partial_j D_1(x-x') [S^i (x),  [S^j (x'), \rho_{\nu} ] ]
.\end{gather}

In the final equation (\ref{A3Final}) we write $K_0$ superoperator, which corresponds to a Hamiltonian that arises due to $\bar{A}_\mu$. The two terms $K_1$ and $K_2$ arise due to entanglement with fluctuations of a  magnetic field. They are real and have a form of a dissipative terms in the Lindblad's equation (see for instance \cite{Lindblad} and \cite{Gorini}).


The derived equation (\ref{A3Final}) describes the effect of neutrino quantum decoherence  due to entanglement with magnetic field fluctuations. In contrast to the approach based on Lindblad's equation there are no free parameters in equation (\ref{A3Final}). The developed approach is more appropriate than the one based on Lindblad's equation for using in studies of neutrino propagation in astrophysical environments with magnetic fields, for example supernovae, because the obtained up to now experimental data on supernovae neutrinos is quite poor in respect, for instance, with the prersent data on solar neutrinos.

\section{Acknowledgments}

The authors are thankful to Konstantin Kouzakov, Alexey Lokhov and Alexander Grigoriev for useful discussions on the topic of the paper.

This work was supported by the Russian Foundation for Basic Research under grants
No.~16-02-01023\,A and No.~17-52-53133\,GFEN\_a.

\end{document}